\documentclass[aps,prl,groupedaddress,twocolumn]{revtex4}
\usepackage{graphicx}
\begin{document} 

\title{High Resolution Imaging of Single Atoms in a Quantum Gas}
\author{Tatjana Gericke}
\author{Peter W\"urtz}
\author{Daniel Reitz}
\author{Tim Langen}
\author{Herwig Ott}
\email[To whom correspondence should be addressed\\E-mail: ]{ott@uni-mainz.de}
\affiliation{Institut f\"ur Physik, Johannes Gutenberg-Universit\"at, 55099 Mainz, Germany}

\date{\today}

\maketitle

\textbf{Our knowledge on ultracold quantum gases is strongly influenced by our ability to probe these objects \cite{andrews1996,schellekens2005,oettl2005,foelling2005}. \textit{In situ} imaging combined with single atom sensitivity is an especially appealing scenario as it can provide direct information on the structure and the correlations of such systems \cite{naraschewski1999,kheruntsyan2005,sykes2008}. For a precise characterization a high spatial resolution is mandatory. In particular, the perspective to study quantum gases in optical lattices \cite{greiner2002,mandel2003,bloch2008} makes a resolution well below one micrometer highly desirable. Here, we report on a novel microscopy technique which is based on scanning electron microscopy and allows for the detection of single atoms inside a quantum gas with a spatial resolution of better than 150 nm. Imaging a Bose-Einstein condensate in a one-dimensional optical lattice with 600 nm period we demonstrate single site addressability in a sub-$\mu$m optical lattice. The technique offers exciting possibilities for the preparation, manipulation and analysis of quantum gases.}

Ultracold atoms can be visualized by various techniques. Absorption imaging \cite{ketterle1999} is the workhorse in most experiments and is typically applied in time of flight in order to increase the cloud size and reduce the optical density. While phase contrast imaging \cite{andrews1996,shin2006} is well suited for trapped quantum gases, fluorescence imaging \cite{hu1994,schlosser2001,teper2006,kuhr2001,schrader2004,nelson2007} is especially attractive as it allows for single atom detection with almost 100\% efficiency. It has been applied to isolated thermal atoms at low densities but has not yet been extended to single atom detection in quantum gases. The best achievable resolution of these optical techniques is ultimately limited by half the wavelength of the used light field - in practice, the best reported resolution is about 1 $\mu$m \cite{schlosser2001}. Direct particle detection of metastable atoms in time of flight \cite{schellekens2005,jeltes2007} and outcoupling of single atoms from a condensate with a radio frequency field \cite{oettl2005} are alternative techniques which have been developed. However, they either cannot be applied to trapped samples \cite{schellekens2005,jeltes2007} or are restricted to one spatial dimension \cite{oettl2005}. Whereas each of those techniques has its specific advantages and applications, a versatile \textit{in situ} detection of single atoms in a quantum gas is lacking. Moreover, a spatial resolution of below 1 $\mu$m which opens the intriguing perspective to resolve single sites in a sub-$\mu$m optical lattice has not yet been achieved.

\begin{figure}[ht]
\includegraphics[scale=0.8]{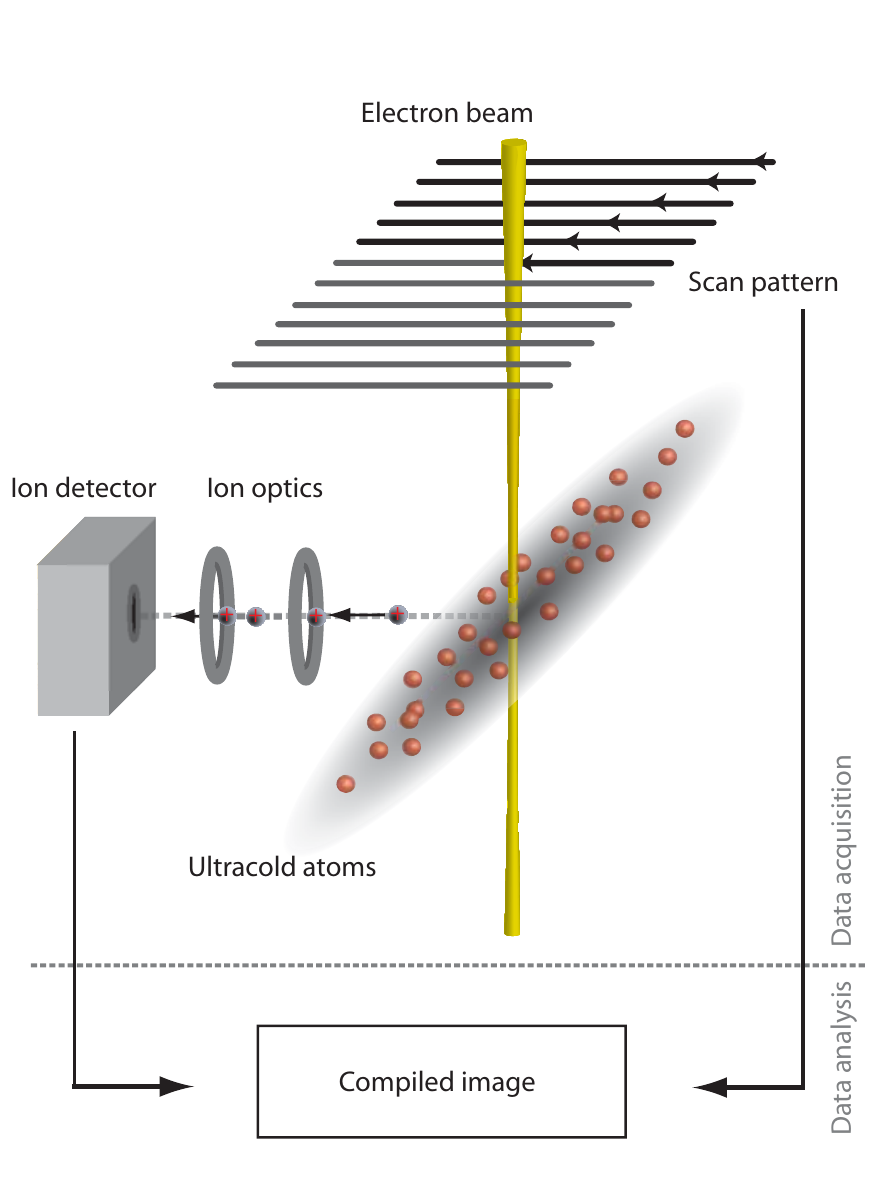}
\caption{Working principle. The atomic ensemble is prepared in an optical dipole trap. An electron beam with variable beam current and diameter is scanned across the cloud. Electron impact ionization produces ions which are guided with an ion optical system towards a channeltron detector. The ion signal together with the scan pattern is used to compile the image.}
\end{figure}

In our experiment we have transferred the principles of scanning electron microscopy to the detection of ultracold atoms (Fig.\,1). A focused electron beam with $6\,$keV electron energy, a full width half maximum (FWHM) diameter of 100-150 nm and a current of 10-20 nA is scanned across a Bose-Einstein condensate of rubidium atoms which is prepared in an optical dipole trap \cite{gericke2007}. The atoms are ionized by electron impact ionization, extracted with an electrostatic field and subsequently detected by an ion detector. The small diameter of the electron beam ensures a high spatial resolution, whereas the ion detection provides single atom sensitivity. The total ionization cross section at $6\,$keV electron energy for rubidium is $\sigma_{\rm{ion}}=3.5\times10^{-17}\textrm{cm}^2$ \cite{bartlett2004} and represents 40\% of all scattering events \cite{schappe1995,schappe1996}. Elastic and inelastic electron-atom collisions constitute the remaining events and lead to atom loss with no detectable signal. As the cross section is eight orders of magnitude smaller compared to the absorption cross section of a resonant photon, the atomic cloud is optically thin for the electron beam. For typical parameters, only 1 out of 500,000 incident electrons undergoes a collision. 

\begin{figure}[ht]
\includegraphics[scale=0.8]{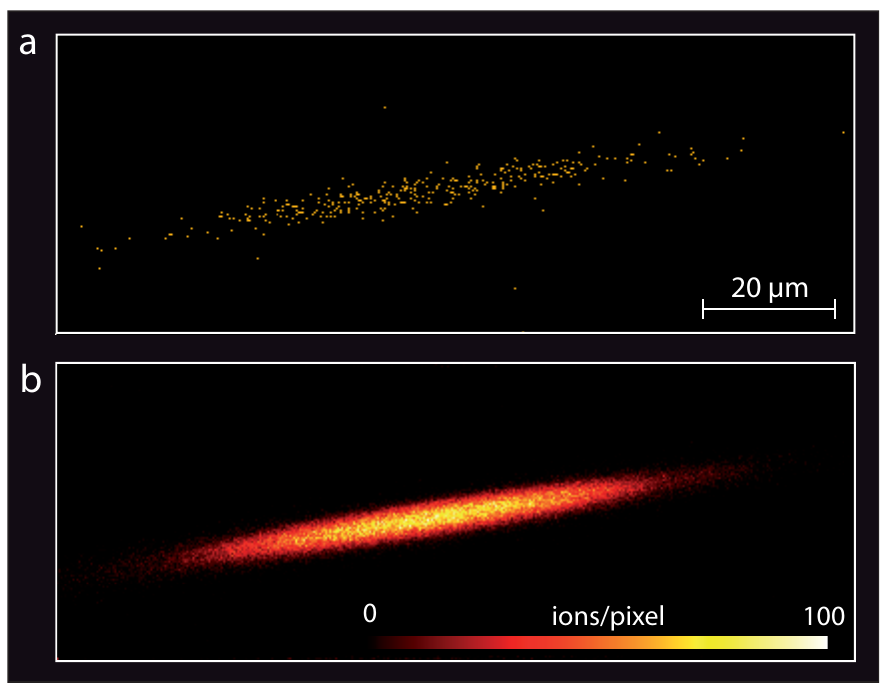}
\caption{Electron microscope image of a trapped Bose-Einstein condensate. In {\bf a} we show an image of a $^{87}$Rb condensate. The image has 400 x 150 pixels with a pixel size of $300\,\textrm{nm}$ x $300\,\textrm{nm}$. Each pixel was illuminated for $2\,\mu\textrm{s}$ with the electron beam ($140\,\textrm{nm}$ FWHM beam diameter). Every dot corresponds to a detected atom. In total, 350 ions were collected during the exposure. The condensate contains about $10^5$ atoms. The sum over 300 images is presented in {\bf b}. Each image was taken in a separate experimental run. }
\end{figure}

When exposed to the electron beam, the probability for the detection of an atom at a position $\{x,y\}$ is given by (see Methods section)
\begin{equation}\label{eq1}
P(x,y) = \frac{I}{e} \sigma_{\rm{ion}} \times \Delta t \times \eta_{\rm{det}} \times N \int dz \, |\phi(x,y,z)|^2.
\end{equation}
Here, $I$ is the electron beam current,  $e$ is the electron charge, $\Delta t$  is the pixel dwell time of the electron beam, $\eta_{\rm{det}}$ is the detector efficiency, $\int dz \, |\phi(x,y,z)|^2$ denotes the column density of the atom's wave function along the propagation direction of the electron beam (z-direction) and $N$ is the number of atoms in the single particle state. In Fig.\,2a we show a scanning electron microscope image of a Bose-Einstein condensate. For our experimental parameters, a fraction of 350 atoms is detected (the total number of atoms in the condensate is about 100,000). In a Bose-Einstein condensate all atoms occupy the same quantum state and the many-body wave function $\psi$ separates into the product of $N$ identical single particle wave functions $\psi(\vec{x}_1,...,\vec{x}_N)=\prod_{i=1}^N\phi(\vec{x}_i)$ , with $N$ being the number of atoms in the condensate. Therefore, the interpretation of the image involves quantum-mechanical concepts: As the single particle wave function $\phi$ extends over the whole atomic cloud, the spatially resolved detection of an atom must be understood as a projective measurement in position space. As a consequence, the retrieved image is intrinsically probabilistic. This is in contrast to almost all microscopy images showing the distribution of individual atoms as in these cases the location of the atoms is already fixed prior to their detection. Another important aspect is related to the Heisenberg uncertainty principle. During the detection process, the atom is coupled to a probe (in our case an electron beam) and energy as well as momentum can be exchanged between them. Consequently, the localization of an atom within a range $\Delta x$ enforces a momentum spread of $\Delta p \geq \hbar/(2\Delta x)$. If $\Delta x$ is smaller than the extension of the wave function $\phi$, substantial momentum transfer is unavoidable and the detected atom is no longer part of the condensate, regardless of the specific experimental realization. Hence, the ionization of the atoms in our scheme does not constitute a serious limitation or drawback. It is even advantageous because it helps rapidly extracting the reaction products from the remaining system, keeping possible perturbations small.

Whether the image in Fig.\,2a is indeed a probabilistic selection of the full atomic distribution according to equation (\ref{eq1}), can be checked by summing over many images (Fig.\,2b) and comparing them to a theoretical density profile. The profile is derived from the so-called semi-ideal model \cite{minguzzi1997,naraschewski1998,gerbier2004}, which describes a bimodal distribution at finite temperature. While the condensate part is obtained from a numerical solution of the 3D-Gross-Pitaevskii equation, the thermal component is modelled as a non-interacting gas in an effective potential, taking into account the repulsion of the thermal atoms by the condensed atoms (see Methods section). The comparison with our data (Fig.\,3a,b) exhibits very good agreement over the whole extension of the cloud including the wings of thermal atoms. This gives not only indirect evidence of the repulsion between the condensate fraction and the thermal component in the trap (Fig.\,3c), but also confirms that the image shown in Fig.\,2a displays a probabilistic selection of atoms.

\begin{figure}[ht]
\includegraphics[scale=0.9]{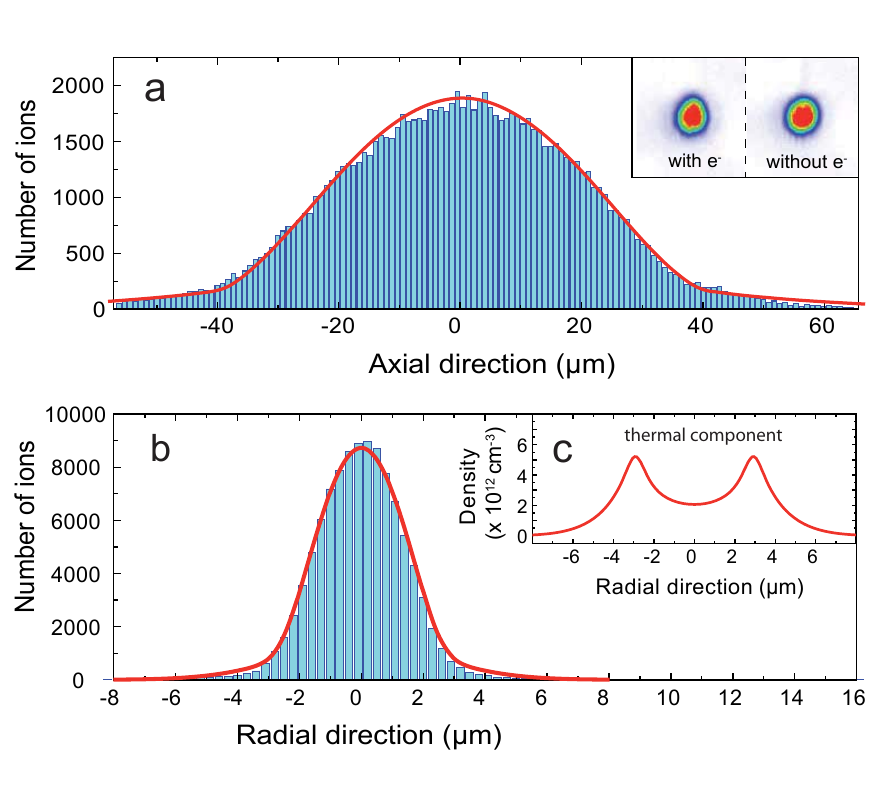}
\caption{Analysis of the Bose-Einstein condensate. In {\bf a} we present the axial distribution of the condensate shown in Fig.\,{2\bf b}, integrated in the radial direction. The experimental data (blue columns) are compared to a bimodal distribution (red line) calculated in the semi-ideal model for a total atom number of $N=115,000$ and a temperature of $T=80\,\textrm{nK}$, corresponding to a condensate fraction of 80\%. The inset shows absorption images of the condensate after $15\,\textrm{ms}$ time of flight with and without exposure to the electron beam. The number of atoms after exposure is reduced by 7\% in average. The distribution in the radial direction is shown in {\bf b}. In {\bf c} we have plotted the radial density of the thermal component in the trap centre as calculated from the model. The minimum is due to the repulsion from the condensate fraction.}
\end{figure}

Comparing the condensate with and without exposure to the electron beam (absorption images in the inset of Fig.\,3a) we do not find any significant difference apart from a reduction in atom number by about 7\%. These losses are composed of two contributions: primary electron atom collisions and secondary collisions of the primary reaction products. We find that every scattered atom or produced ion kicks off on average one more atom. In all these collisions the energy transfer is much larger than the depth of the optical potential and all scattered particles can escape from the trap. Essentially no energy is deposited in the cloud as we observe an additional heating of merely 5 nK after exposure to the electron beam. Thus, the perturbation caused by the detection process is very small. If not, the scanning speed could be made larger than the speed of sound in the condensate providing an effectively unperturbed cloud during the whole imaging sequence. According to equation (\ref{eq1}), high imaging speed is associated with a reduced signal and a convenient setting of the imaging parameters has to be chosen for each application. Most detected ions are singly charged (80\%) but we also find higher charged states of up to $\textrm{Rb}^{7+}$ resulting from inner shell ionization. Only $1$ out of $50$ detected events is due to background gas ionization or dark counts which results in a high signal to noise ratio as evidenced by Fig.\,2. Taking into account a detector efficiency of 30\% the total efficiency for our detection scheme is currently limited to 12\%. It can be increased by a more efficient ion detector and additional photoionization of inelastically scattered atoms. We estimate that a total detection efficiency of more than 50\% could be feasible. 

\begin{figure}[ht]
\includegraphics[scale=0.8]{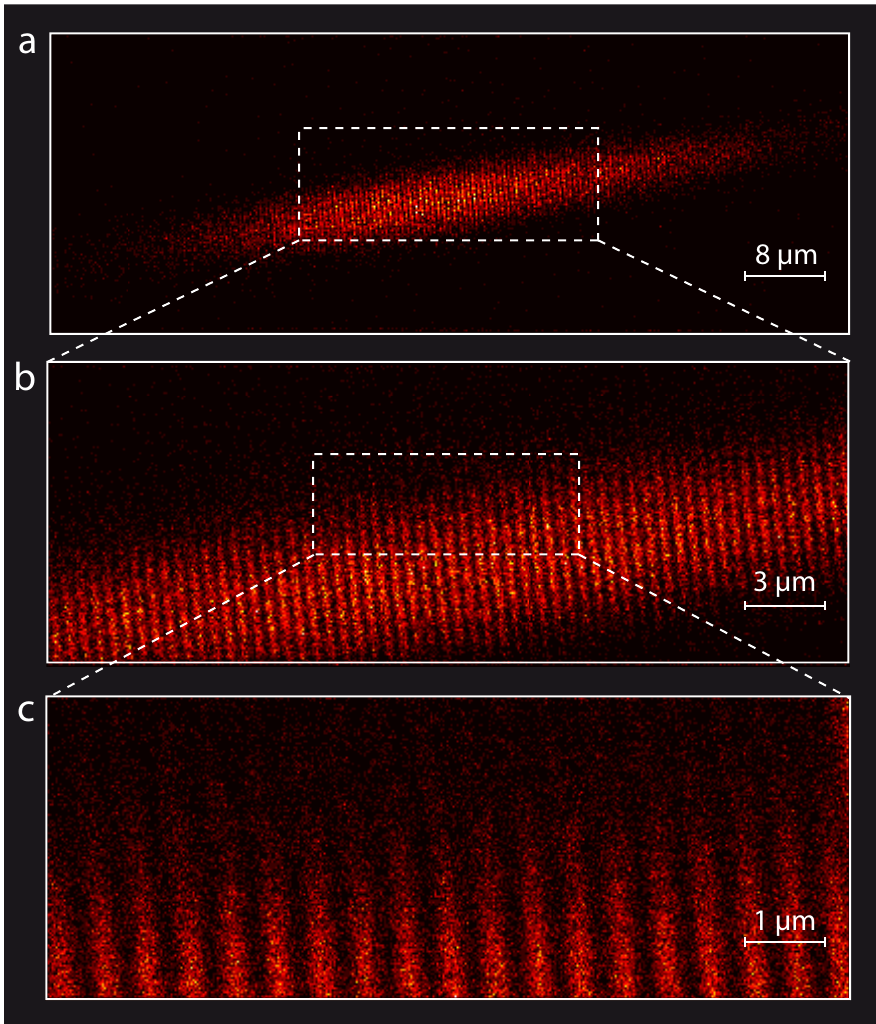}
\caption{Images of a Bose-Einstein condensate loaded in a one-dimensional optical lattice. The lattice which is created by two laser beams ($\lambda=850$\,nm) intersecting under $90^\circ$ has a period of $l=600\,$nm. Each image is the sum of 50 individual images. The pixel size is $200\,\textrm{nm}$ x $200\,\textrm{nm}$ ({\bf a}), $75\,\textrm{nm}$ x $75\,\textrm{nm}$ ({\bf b}), and $25\,\textrm{nm}$ x $25\,\textrm{nm}$ ({\bf c}). The lattice depth was 20 recoil energies $E_r$ ($E_r=\pi^2\hbar^2/(2ml^2)$, with $m$ being the rubidium mass) and the FWHM diameter of the electron beam was $95\,\textrm{nm}$.}
\end{figure}

In order to characterize the resolution of our imaging technique we have loaded the condensate in a one-dimensional optical lattice with $600\,\textrm{nm}$ lattice period. A sequence of electron microscope images with increasing resolution is shown in Fig.\,4a-c. The periodic structure of the potential is clearly resolved with high contrast. As a scanning probe technique is used for the image formation, the addressability of individual lattice sites is demonstrated as well. The atomic density in each lattice site is radially symmetric with a diameter of $6\,\mu$m and a thickness of $300\,\textrm{nm}$ and documents the large depth of focus of the electron optical imaging system.
 
\begin{figure}[ht]
\includegraphics[scale=0.8]{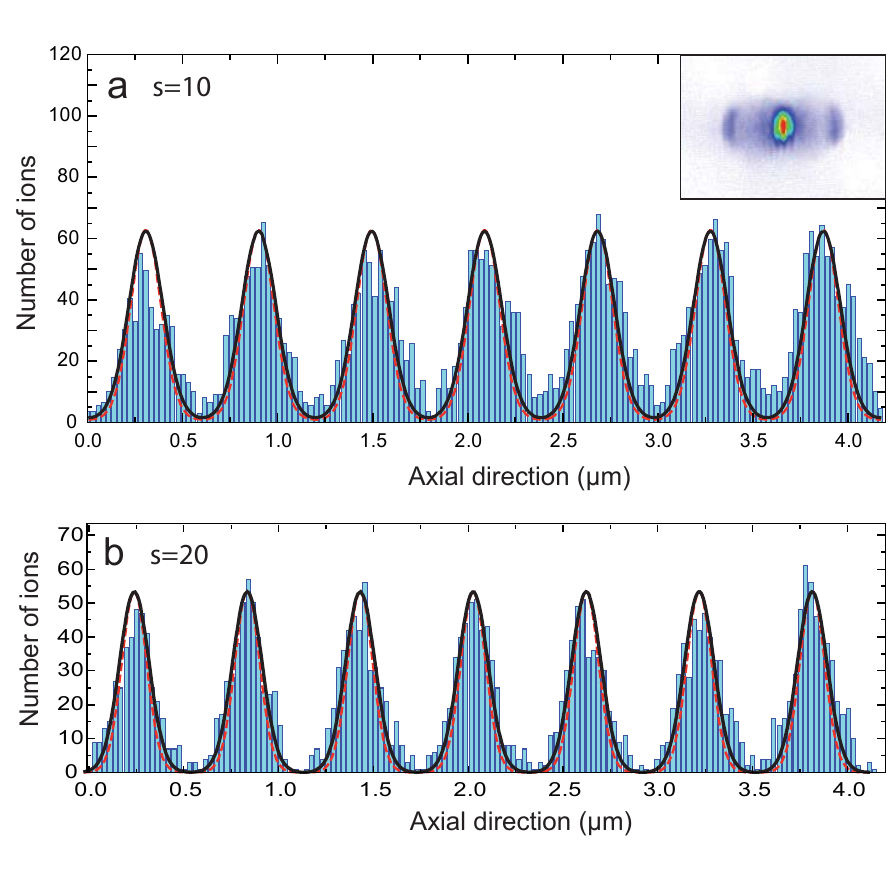}
\caption{Ground state of a Bose-Einstein condensate in a one-dimensional optical lattice. The graphs show integrated line scans for a lattice depth of $s=10$ ({\bf a}) and $s=20$ ({\bf b}), where $s$ measures the lattice depth in units of the recoil energy. The blue columns are the experimental data which are compared to a theoretical model (black line) which is based on the ground state in the lattice potential (red dashed line) convolved with a Gaussian electron beam profile with $95\,\textrm{nm}$ FWHM diameter. An absorption image after 15 ms time of flight (inset in {\bf a}) reveals that the phase coherence of the condensate is preserved after exposure to the electron beam.}
\end{figure}

One of the most intriguing properties of a Bose-Einstein condensate is its macroscopic phase coherence. In a periodic potential the phase coherence can be easily verified by interference experiments. An absorption image of the condensate after a ballistic expansion of $15\,\textrm{ms}$ is shown in the inset of Fig.\,5a. The image was taken after illumination with the electron beam and the appearance of the characteristic diffraction peaks demonstrates that the partial measurement of a subset of atoms does not destroy the coherence of the remaining system. Furthermore, it is an example for a complementary measurement in position and momentum space on a single many-body quantum system. For a quantitative analysis we compare the integrated linescans with the Bloch wave function that describes the ground state of non-interacting atoms in the lattice potential (Fig.\,5a,b). The periodic structure and the shape of the individual on-site wave function are well reproduced for both data sets. Together with the observed interference pattern both, the density distribution and the quantum-mechanical phase are determined and thus, the Bloch wave function is fully characterized. Eventually, we conclude from the good agreement that our imaging technique achieves a spatial resolution of better than $150\,\textrm{nm}$ (see Methods section).

The combination of high spatial resolution and single atom sensitivity will open up new possibilities for the \textit{in situ} study of spatial and temporal correlations in trapped quantum gases\cite{naraschewski1999,kheruntsyan2005,sykes2008}. Previous experimental work on correlations in trapped \cite{esteve2006} and expanding \cite{jeltes2007,foelling2005,greiner2005} gases has already demonstrated the high potential of such measurements. In optical lattice systems, the technique cannot only be used as a powerful characterization tool. Removing atoms from specific lattice sites, it also allows for the preparation of tailored quantum systems. Taking advantage of the electron beam's magnetic field even a coherent manipulation of single atoms could be feasible.
\\
\\
\\
\textbf{Methods}
\\
\\
\textbf{Detection probability:}
An atom that is located in the centre of a Gaussian electron beam with a radial current density of $j(\rho) = j_0 \exp(-\rho^2 / 2 \rho_0^2)$, has a lifetime against electron impact of $\tau = e / (j_0 \sigma_{\rm{tot}})$. Here, $j_0=I/(2 \pi \rho_0^2)$ is the current density in the beam centre, $I$ is the beam current, $\rho_0$ is the $\sigma$-width of the beam, $e$ is the electron charge, and $\sigma_{\rm{tot}} = 9\times10^{-17}$ cm$^2$ is the total electron scattering cross section for rubidium at 6 keV electron energy. For typical beam parameters ($I=23$ nA, $\textrm{FWHM}=140$ nm, corresponding to $\rho_0=85$ nm) we obtain $\tau \approx 17 \,\mu$s. If the pixel dwell time $\Delta t$ is much smaller than $\tau$, the probability  for a scattering event (ionization, elastic or inelastic scattering) is given by
$$ w = 1 - e^{-\frac{\Delta t}{\tau}} \approx \frac{\Delta t}{\tau} = \frac{j_0}{e} \sigma_{\rm{tot}} \times \Delta t. $$

If the atom is described by a wave function $\phi(x,y,z)$ and if we assume that the beam is much smaller than the extension of the wave function, the probability of a scattering event at the position $\{x,y\}$ is given by
\begin{equation}
w(x,y) = \frac{I}{e} \sigma_{\rm{tot}} \times \Delta t \times \int dz \, | \phi(x,y,z) |^2.
\end{equation}
Multiplying equation (2) with the ion production efficiency $\sigma_{\rm{ion}}/\sigma_{\rm{tot}}$, the detector efficiency $\eta_{\rm{det}}$, and the total number of atoms gives equation (1) of the main text.

\textbf{Bimodal distribution:}
For a given number of condensed atoms $N_0$ we numerically solve the Gross-Pitaevskii equation 
\begin{equation}
\left[ -\frac{\hbar^2}{2m} \vec{\nabla} + V_{\rm{ext}}(\vec{r}) + g |\psi(\vec{r})|^2 \right] \psi(\vec{r}) = \mu \psi(\vec{r})
\end{equation}
using an imaginary time propagation algorithm. The external potential is cylindrically symmetric and has the form $V_{\rm{ext}}(\rho,z) = \frac{1}{2} m ( \omega_\rho^2 \rho^2 + \omega_a^2 z^2)$, where $z$ denotes the axial direction of the condensate, $\omega_a = 2\pi \times 12$ Hz ($\omega_\rho = 2\pi \times 170$ Hz) is the axial (radial) oscillation frequency of the dipole trap, $\mu$ is the chemical potential, $g = 4 \pi \hbar^2 a / m$ is the coupling constant, and $m$ is the rubidium mass. For the s-wave scattering length we use a value of $a = 101\,a_0$, with $a_0$ being the Bohr radius. In our experiment we produce a spinor condensate in the $|F=1\rangle$ ground state of rubidium. For the model presented here we neglect the spinor nature because the difference in the scattering lengths for the $F=0$ and $F=2$ scattering channels are only 1\% \cite{vankempen2002}. The condensate wave function is normalized to the total number of condensed atoms, $N_0 = \int d^3x \, |\psi(\vec{r})|^2$. The numerical solution of equation (3) is used to model an effective potential for the thermal component
$$ V_{\rm{eff}}(\vec{r}) = V_{\rm{ext}}(\vec{r}) + 2 g |\psi(\vec{r})|^2 .$$
The density distribution of the thermal component is then given by 
\begin{equation}
n_{\rm{th}}(\vec{r}) = \lambda_{\rm{th}}^{-3} g_{3/2}(z)
\end{equation}
with a modified fugacity
$$z = \exp\left[-(V_{\rm{eff}}(\vec{r}) - \mu) / k_B T\right].$$
Here, $\lambda_{\rm{th}} = \sqrt{2\pi \hbar^2 / m k_B T}$ is the thermal de-Broglie wavelength, $k_B$ is the Boltzmann constant and $T$ is the temperature. The number of atoms in the thermal component is given by $N_{\rm{th}} = \int d^3x \, n_{\rm{th}}(\vec{r})$ and the total number of atoms is $N = N_0 + N_{\rm{th}}$.

\textbf{Spatial resolution:}
The size of the electron beam can be determined independently by scanning the beam across a sharp edge of a movable test target which is implemented in the vacuum chamber. We define the resolution as the distance between two neighbouring point-like scatterers where the signal intensity in between drops to 75\%. This definition is the analogue to the Rayleigh criterion in optics and for our system translates into a resolution of $d = 1.18 \times \textrm{FWHM}$, assuming a Gaussian beam profile. The electron beam used for the measurement in Fig. 4 and 5 of the main text has a diameter of 95 nm FWHM, corresponding to a resolution of 115 nm. The good agreement between the experimental line scan and the theoretical model in Fig. 5 proves that a similar resolution is achieved for the electron microscope images of ultracold atoms.

We would like to thank A. Widera and T. Best for valuable discussions and C. Utfeld for contributions in the early stage of the experiment. This work was funded through the DFG and the Forschungsfond of the University of Mainz.

\end{document}